# Observation of possible excitonic charge density waves and metal-insulator transitions in atomically thin semimetals

Qiang Gao[1]*, Yang-hao Chan[2,3]*, Pengfei Jiao[1]*, Haiyang Chen[1], Shuaishuai Yin[4], Kanjanaporn Tangprapha[2,5], Yichen Yang[6], Xiaolong Li[4], Zhengtai Liu[6], Dawei Shen[6,7], Shengwei Jiang[1], Peng Chen[1]

[1]Key Laboratory of Artificial Structures and Quantum Control (Ministry of Education), Shenyang National Laboratory for Materials Science, Shanghai Center for Complex Physics, School of Physics and Astronomy, Shanghai Jiao Tong University, Shanghai 200240, China.

[2]Institute of Atomic and Molecular Sciences, Academia Sinica, Taipei 10617, Taiwan.

[3]Physics Division, National Center for Theoretical Sciences, Taipei 10617, Taiwan.

[4]Shanghai Synchrotron Radiation Facility, Shanghai Advanced Research Institute, Chinese Academy of Sciences, Shanghai 201204, China

[5]College of Engineering, International Advanced Technology Program, National Taiwan University of Science and Technology

[6]State Key Laboratory of Functional Materials for Informatics, Shanghai Institute of Microsystem and Information Technology (SIMIT), Chinese Academy of Sciences, Shanghai 200050, China.

[7]National Synchrotron Radiation Laboratory, University of Science and Technology of China, 42 South Hezuohua Road, Hefei, Anhui 230029, China.

*These authors contributed equally to this work.

**Charge density wave (CDW) is a collective quantum phenomenon with a charge modulation in solids[1-2]. Condensation of electron and hole pairs with finite momentum will lead to such an ordered state[3-7]. However, lattice symmetry breaking manifested as the softening of**

phonon modes can occur simultaneously, which makes it difficult to disentangle the origin of the transition[8-14]. Here, we report a condensed phase in low dimensional $HfTe_2$, whereas angle-resolved photoemission spectroscopy (ARPES) measurements show a metal-insulator transition by lowering the temperature in single triatomic layer (TL) $HfTe_2$. A full gap opening, renormalization of the bands, and emergence of replica bands at the $\bar{M}$ point are observed in the low temperatures, indicating formation of a CDW in the ground state. Raman spectroscopy shows no sign of lattice distortion within the detection limit. The results are corroborated by first-principles calculations, demonstrating the electronic origin of the CDW. By adding more layers, the phase transition is suppressed and completely destroyed at 3 TL because of the increased screening around the Fermi surface. Interestingly, a small amount of electron doping in 1 TL film during the growth significantly raises the transition temperature ($T_C$), which is attributed to a reduced screening effect and a more balanced electron and hole carrier density. Our results indicate a CDW formation mechanism consistent with the excitonic insulator phase in low dimensional $HfTe_2$ and open up opportunity for realization of novel quantum states based on exciton condensation.

In a semimetal, exciton condensation spontaneously occurs at low temperatures when the density of carriers is small enough and the Coulomb interaction between electrons and holes is weakly screened[1-5]. It can give rise to a charge density modulation when the correlated electrons and holes are separated in the momentum space and an insulating phase can be achieved when the Coulomb interaction is strong enough. Reduced screening effect in low dimensional layered materials like transition metal dichalcogenides provides an ideal condition for searching such a state. A number of candidates have been proposed and signatures of the phase transition include the renormalizations of band structure, energy gap opening, and so on[8-20]. However, similar features can be observed in a lattice instability driven structure transition[9,14], which makes it challenging to exclusively identify whether the transition mechanism lies in the electronic or structural origin as they are coupled with each other through the electron-phonon interaction. In this regard, observation of a metal-insulator transition without phonon softening is a key evidence on an excitonic insulator phase.

Here, we study $HfTe_2$, a semimetal with an indirect band overlap of about 0.2-0.3 eV and no phase transition was observed in the bulk form when lowering the temperature[21-24]. We have successfully grown high-quality single-layer and multi-layer $HfTe_2$ films using molecular beam epitaxy (MBE). Atomic layer-by-layer growth was monitored via reflective high energy electron diffraction (RHEED), as shown in Extended Data Fig. 1b, d. Unlike excitonic insulators realized in spatially separated double layers[25-29], a problem in a natural solid is the controllability of the exciton population. This difficulty can be overcome by introducing free carriers in the films during the growth through substrates with different doping levels and the chemical potential of the electron and hole will be modified accordingly. ARPES measurements with both lab-based and synchrotron-based setups are used to characterize the band structures of $HfTe_2$ thin films with varied photon energies.

Fig. 1 shows ARPES spectra taken along the $\overline{\Gamma M}$ direction for multilayer $HfTe_2$ thin films with thickness ranging from 1 to 3.5 triatomic layers (TLs) grown on highly oriented pyrolytic graphite (HOPG) substrates. The measured data at 250 K (Fig. 1a) are in excellent agreement with theoretical results for the normal phase[30]. We obtain the number of layers according to the RHEED intensity oscillation (Extended Data Fig. 1b), and the position/intensity of subbands from band multiplication effect. For the 1 TL film, two hole-like valence bands centered about $\overline{\Gamma}$ are Te-5$p$-derived and no conduction bands derived from Hf 5$d$ states are observed, as shown in Fig.

$1a^{31}$. The sample is metallic as the topmost valence band crossing the Fermi level at 250 K. The resulting metallic Fermi surface do not have regions suitable for nesting and hence one can rule out the possibility of a nesting induced CDW. The topmost valence band develops a flat top with decreasing temperature and a full gap is opening (Fig. 1b). These results indicate the system becomes an insulator/semiconductor at low temperatures. Importantly, the 10 K data show additionally a weak replica of the topmost valence band shifted from $\bar{\Gamma}$ to $\bar{M}$, a spectroscopic signature of a $(2 \times 2)$ band folding and CDW formation. Note that there is no replica for the lower valence band at -0.57 eV since it resides far away from the Fermi level due to the large splitting caused by the strong spin-orbital coupling in Te and the interaction between conduction and the lower valence band is too weak.

As the film thickness increases, the band structure shows layer-by-layer evolution (Fig. 1a). When the thickness is above 2 TL, the conduction band edge becomes visible, indicating an increment of the carrier density around the Fermi level in multilayer films. At 10 K, gaps opening around the $\bar{\Gamma}$ point and backfolded bands at the $\bar{M}$ point are also observed for $N < 3$ TL, demonstrating the CDW state persists in low dimensional $HfTe_2$. This behavior is different from other CDW materials like $TiSe_2$ and $ZrTe_2$, in which, either CDW survives from single layer to the bulk or only in the single-layer film[15,32]. The differences manifest in different structural instability and band structure, as demonstrated in below.

We performed Raman spectroscopy measurements to probe the phonon response with the phase transition. The temperature dependent Raman scattering results on 1 TL $HfTe_2$ films are shown in Fig. 1d. Four Raman peaks at 2.9, 3.3, 3.8 and 4.4 THz (2E + 2A1) revealed at all measured temperatures agree well with the calculations in Fig. 1f and those from the bulk[33]. The lowest measured temperature (1.6 K) is well below the $T_C$ and no significant changes associated with a typical CDW transition including softening of phonon modes and emergence of CDW amplitude modes were observed across the transition[34,35]. The results indicate no obvious lattice distortion in the ordered state and the observed CDW by ARPES is in principle electronic. More Raman data on 1 TL $HfTe_2$ grown with a different substrate show consistent results (Extended Data Fig. 9). Synchrotron x-ray results, as another probe on structural change, show no sign of half-order peaks confirming the lattice distortion is quite small in 1 TL $HfTe_2$ (Fig. 1e).

We computed the phonon dispersion for $N$-layer $HfTe_2$ ($N = 1, 2, 3$) in their $(1 \times 1)$ structure from first-principles, as shown in Fig. 1f and Extended Data Fig. 8. The results show small dips in acoustic branches at the $\bar{M}$ point which might suggest a tendency for $(2 \times 2)$ distortion, but there are no imaginary frequencies that would correspond to the structural instabilities. Of note is that electronic interaction will also affect the lattice as a first order effect and cause such a distortion tendency. Additional phonon calculations were performed at the slightly different experimental lattice constant and we do not find a $(2 \times 2)$ structural transition within our numerical accuracy and the functional investigated.

The observed spectra features including strong valence band renormalization, emergence of backfolded bands, and the metal-insulator phase transition are consistent with the excitonic insulator scenario that excitons with finite momentum condense in low dimensional $HfTe_2$. The evolution of the CDW state in multilayer films can be understood as a raise of the carrier density in the thicker samples, which results in a stronger screening effect, an unfavorable condition for the formation of bound electron-hole state. We extract the carrier density from the Luttinger area of the Fermi surface in the normal phase[36]. As an example, the carrier density for a 1 TL film is ~7.0 x $10^{12}$/cm$^2$ (hole density $p$) only at the $\bar{\Gamma}$ point. It increases to ~4.8 x $10^{12}$/cm$^2$ at the $\bar{\Gamma}$ point plus ~2.4 x $10^{13}$/cm$^2$ (electron density $n$) at the $\bar{M}$ point in 3 TL and the features of exciton condensation disappear at this thickness.

In the normal phase, the Fermi surface is dominated by a hole-like pocket in 1 TL HfTe$_2$/HOPG and the conduction bands are located above the Fermi level, which implies a carrier density imbalance ($n$ - $p$). Electron doping can modify the chemical potential of the system, which is a promising way to investigate the CDW mechanism. A wide range of $n$-type doping were achieved by growing the films on epitaxial bilayer graphene (BLG) on SiC with varied resistances. Fig. 2, a-e, present the ARPES spectra along the $\overline{\Gamma M}$ direction for 1 TL HfTe$_2$ with different doping levels. The valence bands shift down away from the Fermi level upon doping (Fig. 2b) and a conduction band appears with further doping so that the electron density dominates the Fermi surface (Fig. 2d, e). At 10 K, the gap opening and backfolded band corresponding to the CDW state are also observed for electron doped systems and conduction band edge emerges at the $\overline{M}$ point as a consequence of $n$-type doping.

We investigate the metal-insulator transition by taking systematic scans of the bands along $\overline{\Gamma M}$ (Fig. 3) as a function of temperature. Selected temperature dependent ARPES spectra for a doped sample are shown in Extended Data Fig. 2. The energy distribution curves (EDCs) at the $\overline{\Gamma}$ point shown in Fig. 3a, d were symmetrized with respect to the Fermi level to illustrate the gap formation (Fig. 3b, e). At low temperatures, the symmetrized EDCs show the valley structures at the Fermi level, indicating opening of a gap around the zone center. At high temperatures, a peak at the Fermi level indicates the valence band crosses over the Fermi level and the gap is zero. The energy gap for 1 TL HfTe$_2$/HOPG is determined to be 83 meV at 10 K, which is underestimated since the conduction band bottom are above the Fermi level. However, it does not deviate much from the actual value because a small amount of electron doping can bring the conduction edge down below the Fermi level (Fig. 2f, g). The obtained gap value does not vary significantly with small doping (80 meV and 84 meV for HfTe$_2$/BLG/SiC films with a carrier density imbalance of -5.4 x $10^{12}$ and 1.5 x $10^{12}$/cm$^2$, respectively). We extract the transition temperature by fitting to the data using a semi-phenomenological BCS mean-field equation (red curves in Fig. 3c, f)[37]. The results demonstrate $T_C$ in the electron doped film ($n$ - $p$ = -5.4 x $10^{12}$/cm$^2$) is ~2.3 times of that in the sample on HOPG, a quite unusual phenomenon for known typical Peierls-type CDW materials, in which $T_C$ either slightly changes or decreases because of the lattice defect and disorder introduced with doping[38,39]. We also extract the relative backfolded band intensity, defined by the red dashed rectangle shown in Fig. 2f, with respect to the main valence band intensity as a quantitative measure of the folding effect. The integrated intensity as a function of temperature (blue triangles in Fig. 3c,f) agrees well with the trend based on the gap analysis.

The above observed doping-dependent behavior is consistent with the picture of an excitonic insulator induced CDW. As shown in the summarized results in phase diagrams (Fig. 4), the highest $T_C$ is achieved around $n = p$, a consequence of reduced screening effect from decreased hole carrier density and a more balanced electron-hole density favoring exciton formation. Heavy doping suppresses the formation of excitons and leads to a smaller value or zero of $T_C$. We note that $\frac{2\Delta}{k_B T_C}$ ~ 5.5 for a carrier density of $n$ - $p$ = -5.4 x $10^{12}$/cm$^2$, which is larger compared to the typical BCS parameter of 3.52. The value increases to 13.0 for 1 TL HfTe$_2$/HOPG with a carrier density deviating away from the carrier balanced point ($n = p$) and beyond the BCS mean-field theory. The energy gap in 1 TL HfTe$_2$ at 10 K is relatively independent of doping over a doping region (indicated as a shadow area in Fig. 4a) regardless of the large $T_C$ differences (74 - 175 K). Similar behavior has been reported in high temperature superconductors[40,41]. A comprehensive theoretical investigation is needed to unveil the doping dependence of the excitonic transition, which is complicated by the density dependent screening strength.

Thickness dependent phase transition can be understood in terms of carrier density dependence of screening effects. We include the data points from different thicknesses in Fig. 1 according to the estimated carrier density. As the band overlap is increased in multilayer films, the density of states become higher around the Fermi level. The system represents a $p$-type doping in thinner films (N $\leq$ 2) but becomes more $n$-type doped since the conduction band edge shown below the Fermi level when the thickness is above 2 TL. It is not a rigid shift of the chemical potential for both electrons and holes in multilayer films since the overlap between valence and conduction bands becomes larger with increasing thickness. Although carrier density imbalance is reduced in certain thickness (e.g. 2.5 TL), $T_C$ still becomes smaller as the screening effect dominates, which also explains that $T_C$ is suppressed much faster in multilayer films compared with the doped 1 TL $HfTe_2$ (Fig. 4a). As shown in Fig. 4b, $T_C$ in principle decreases with additional total carrier density, $n + p$. As the screening strength is proportional to total carrier density in a simple approximation, the dependence of $T_C$ on $n + p$ demonstrates the important role of the screening effect in the formation of excitonic insulating state in semimetals.

Our experimental and theoretical results thus provide evidence of excitonic insulator in low dimensional $HfTe_2$. Unlike the well-known candidates $TiSe_2$ and recently discovered single-layer $ZrTe_2$, in which lattice distortion is still evidenced in the calculated phonon dispersions although it might not be the main driving force[8,9,15], $HfTe_2$ is unique and an intrinsic excitonic insulator for the electronic character of the transition. Furthermore, the nontrivial topological character for $N \geq 2$ TL systems (Extended Data Fig. 6 and Extended Data Fig. 7) are promising for realization of topological excitonic insulating phase and useful for studying the exotic quantum effects and phenomena raised by correlation between different orders[16].

**Fig. 1 | Thickness dependence of the band structure and phonon dispersions of low dimensional HfTe₂.** ARPES maps along $\overline{\Gamma M}$ for low dimensional HfTe₂ taken with Helium lamp (21.2 eV) at (**a**) 250 K and (**b**) 10 K. (**c**) Corresponding symmetrized ARPES maps in energy about the Fermi level at 10 K, which show gaps for $N < 3$ TL in the condensed phase. (**d**) Temperature-dependent Raman scattering spectra for 1 TL HfTe₂ grown on HOPG, showing no significant changes across the transition temperature. (**e**) Line scans taken along the $\overline{\Gamma M}$ direction in reciprocal space for the 1 TL HfTe₂/BLG/SiC. Integral Bragg peak is shown at $q = -1$ in terms of reciprocal lattice units (r. l. u.) and no half-order peaks are observed. Inset: zoom-in view of the scans around $q$ at -1/2 and -3/2. (**f**) Phonon dispersions for 1 TL HfTe₂ in the normal phase computed from first principles. No imaginary modes are observed.

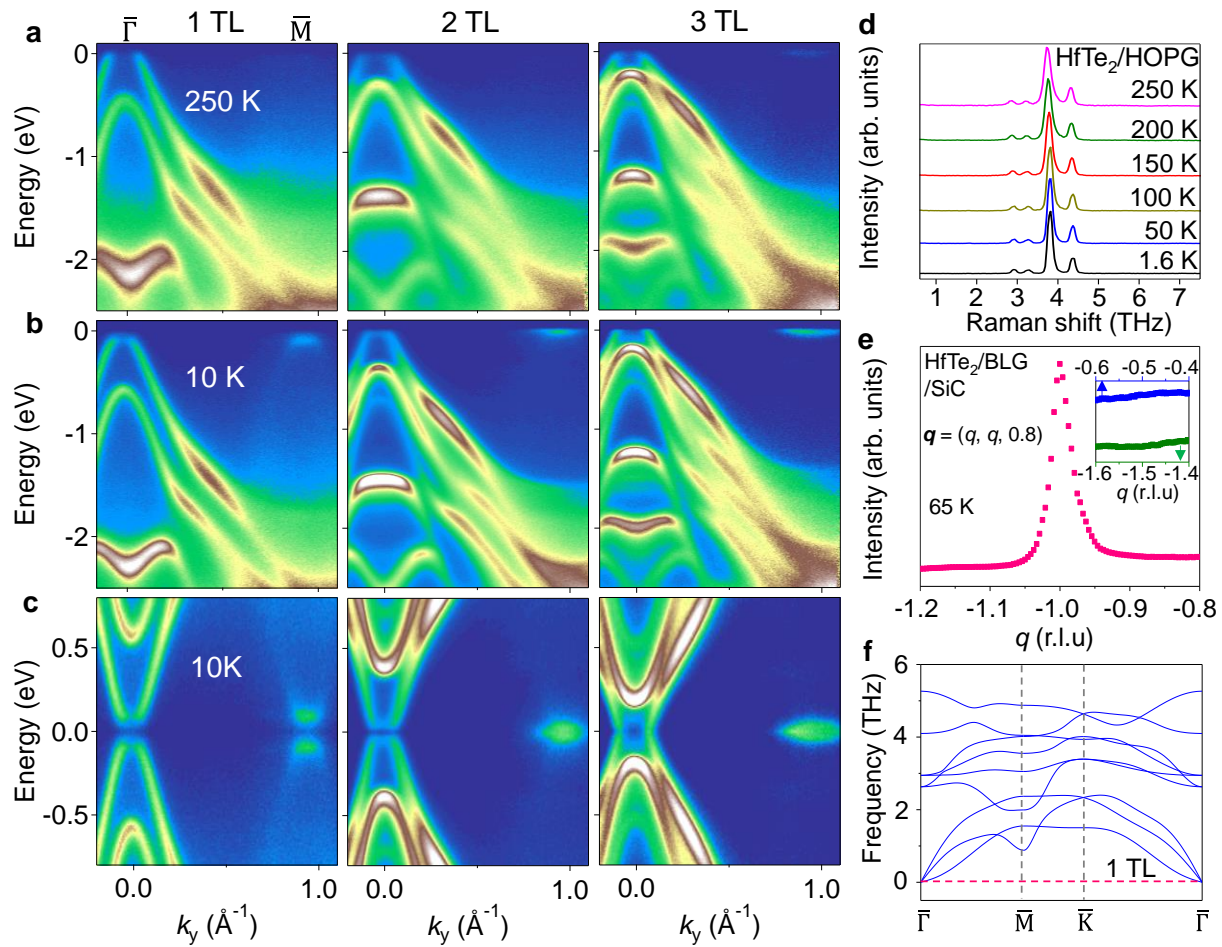

**Fig. 2 | Effect of electron doping on the band structure of 1 TL HfTe₂.** ARPES spectra taken along the $\overline{\Gamma M}$ direction at (**a**)-(**e**) 250K and (**f**)-(**j**) 10 K with different electron doping levels. The data at 10 K show the gap opening around the zone center and folded bands around the $\overline{M}$ point. The red dashed box indicates a region of interest used for integrating the ARPES intensity as a measure of the folded-band intensity. Measured Fermi surface maps for (**k**) HfTe₂/HOPG and (**l**) HfTe₂/BLG/SiC ($n - p$ = -5.4 x $10^{12}$/cm²) at 250 K in the normal phase and 10 K in the condensed phase obtained by integrating the ARPES intensity over ±10 meV about the Fermi level. The Fermi surface contour in the normal phase consists of a hole pocket of circle shape around the zone center in both cases. In the condensed phase, the hole pocket disappears because of the gap opening around the zone center. Electron-hole density imbalance ($n - p$) determined from the Luttinger area of the Fermi surface in the normal phase is shown on top of the panels.

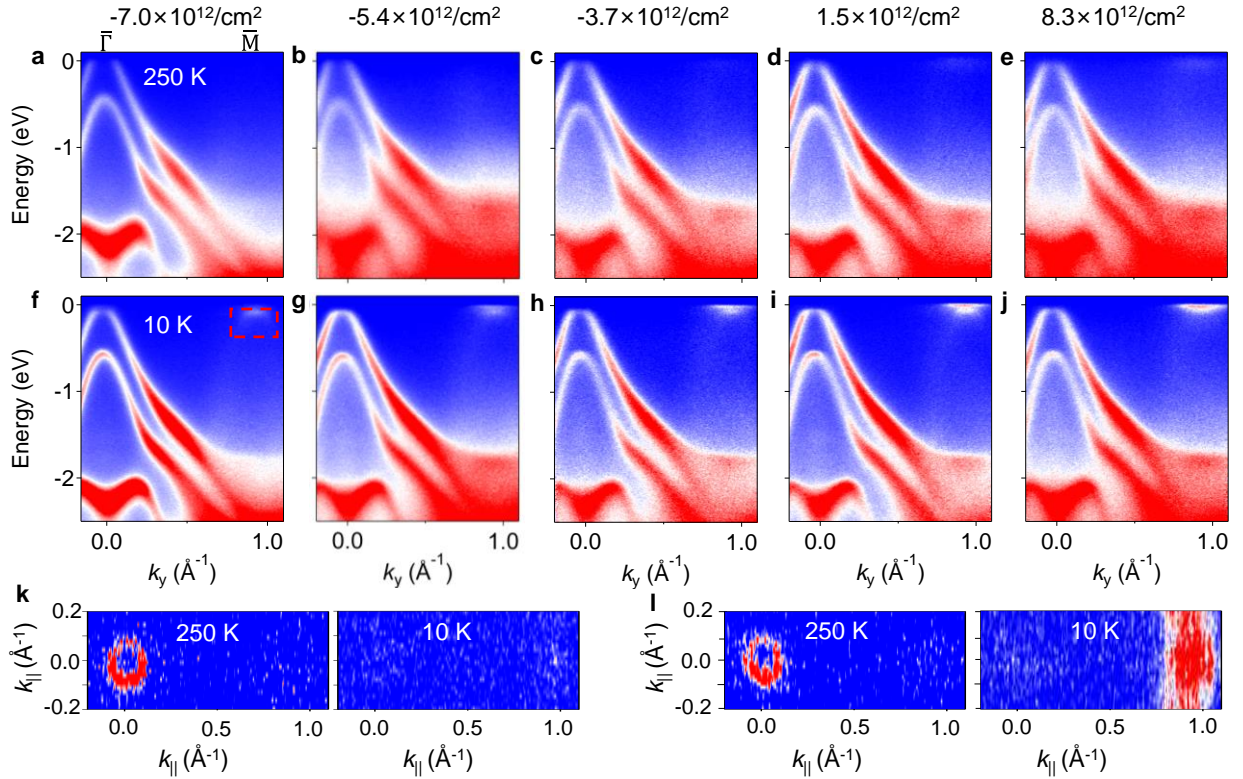

**Fig. 3 | Temperature dependence of the band structure and the energy gaps.** EDCs at the zone center at selected temperatures between 10 and 250 K for doping levels of (**a**) -7.0 x $10^{12}$/cm$^2$ and (**d**) -5.4 x $10^{12}$/cm$^2$, respectively. (**b, e**) Corresponding symmetrized EDCs. The temperature dependence of the Fermi-Dirac distribution can be canceled out by symmetrization. The fits to a phenomenological BCS-type function[37] are indicated as solid curves. (**c, f**) The extracted temperature dependence of the square of the energy gap. The red curves are fitting results using the BCS-type mean-field equation. Relative integrated ARPES intensities over the region of interest (indicated as a red dashed box in Fig. 2f) as a function of temperature are shown as blue triangles.

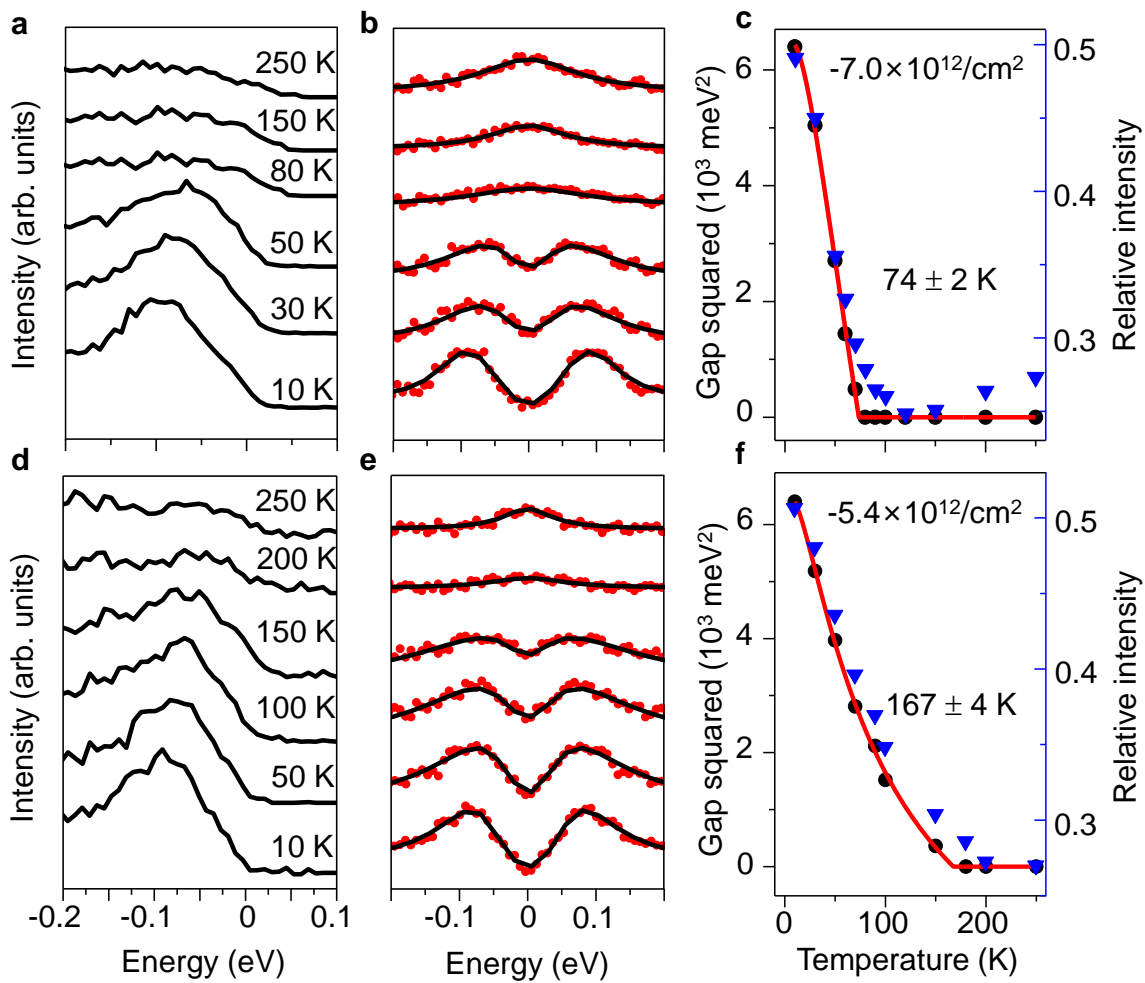

**Fig. 4 | Excitonic insulator phase diagrams. (a)** Transition temperature as a function of electron-hole density imbalance ($n − p$). The red pentagrams denote the $T_C$s from the doped samples and the blue triangles are the data from the multilayer films, which are determined from the temperature dependent ARPES measurements. The shadow area denotes the region where the measured gaps at 10 K are almost the same. **(b)** Transition temperature as a function of electron-hole total density ($n + p$), showing that $T_C$ is suppressed with stronger screening strength.

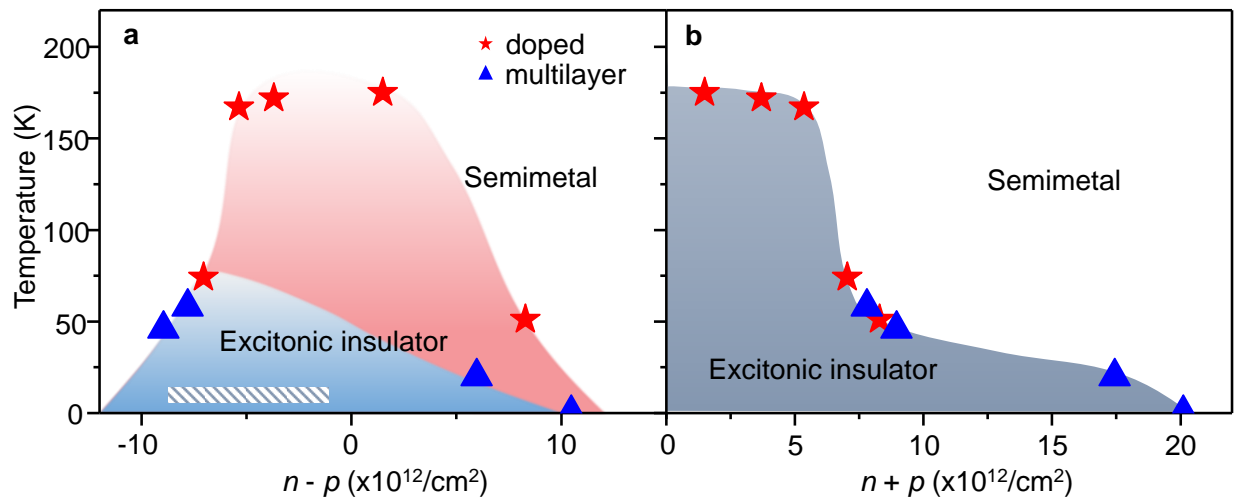

**Methods**

**Film growth**

We grow HfTe$_2$ thin films on substrates of HOPG or SiC in the integrated MBE/ARPES systems at the lab in Shanghai Jiao Tong University. To achieve different electron doping levels, SiC substrates with varied resistivity (0.015-0.1 Ω·cm) were used. SiC were flash-annealed for multiple cycles to form a well-ordered bilayer graphene on the surface. Single-layer and multilayer HfTe$_2$ films were grown on top of the substrate by co-evaporating high purity Hf and Te from an electron-beam evaporator and a Knudsen effusion cell, respectively, while the substrate was maintained at 400 °C. Multilayer films were annealed at 500 °C for 20 minutes after growth to achieve better crystalline quality. The growth process and thickness of the films was monitored by RHEED, and the growth rate was set to 45 minutes per layer of HfTe$_2$.

**ARPES and x-ray measurements**

After growth, the films were transferred *in situ* to the ARPES system at the lab in Shanghai Jiao Tong University or transferred via a vacuum suitcase to beamline 03U at Shanghai Synchrotron Radiation Facility[42], in which the film surface was recovered by annealing at 400 °C before the experiments. ARPES measurements were performed at a base pressure of ~5x10$^{-11}$ mbar with in-laboratory He discharge lamp (He-I 21.2 eV) and 25-100 eV photons at synchrotron using Scienta DA30 analyzers. Energy resolution is better than 15 meV and angular resolution is around 0.2°. Each sample's crystallographic orientation was precisely determined from the symmetry of constant-energy-contour ARPES maps. The Fermi level is determined by fitting ARPES spectra from a polycrystalline gold sample. The carrier density with doping is determined from the Luttinger area of the Fermi surface around the $\bar{\Gamma}$ and $\bar{M}$ points based on $N = g_v \frac{k_F^2}{2\pi}$[43]. As the conduction band does not cross the Fermi level for the highest doping level achieved in this work, the carrier densities will be overestimated using the formula above. We can still extract the carrier density from the conduction band tail enclosed by the Fermi surface which is directly proportional to the particle density given by $N = \sum_k \int_{-\infty}^{E_F} f(\omega) A_k(\omega) d\omega$, where $f$(w) is the Fermi-Dirac function and a Lorentian is used for spectral function $A_k$(w). A simple approximation is to use the ratio of the tail area to that of the conduction band as a scale (Extended Data Fig. 11).

X-ray diffraction measurements were performed on the surface x-ray scattering station (BL02U2) at the shanghai synchrotron radiation facility. The energy of the incident x-ray radiation was chosen to be 9.8 keV. The scattered radiation was detected by a Eiger 500K area detector.

**Raman measurements**

Non-polarized Raman spectra were performed under normal incidence with a HeNe laser centred at 633.1 nm. The reflected radiation was analyzed with a grating spectrometer equipped with a liquid-nitrogen-cooled charge coupled device. A combination of one reflective Bragg grating and two Bragg notch filters removed the majority of the laser side bands and allowed measurements of the Raman shift down to ~5 cm$^{-1}$. The laser power was kept below 70 μW to prevent significant laser heating of the samples. To minimize sample degradation, the films were capped with thin layers (~6 nm) of amorphous Te after the growth and transferred to the Raman cryostat in minutes.

With (2 x 2) distortion, phonon modes at the $\bar{M}$ point will fold over to the $\bar{\Gamma}$ point. However, no new peaks emerge in the Raman data at low temperatures, indicating the intensities of these new modes are very weak and the structural distortion is small. A CDW in general will have collective excitations like amplitude modes. As Raman directly probes the polarizability change, the measured intensity is proportional to this polarizability change which includes collective excitations and lattice distortions. The collective excitations should be a weak effect as the intensity of the folded band in $HfTe_2$ is quite weak compared to the other candidates like $TiSe_2$. Moreover, our Raman and x-ray results show no obvious structural change in the system, indicating a tiny lattice distortion. Although a very small Raman intensity of collective excitations could be expected from a CDW driven lattice symmetry breaking, it will be well beyond the detection limit of the current methods.

## Computational details of density functional theory from first principles

Calculations were performed using the Vienna ab initio package (VASP)[44,45] with the projector augmented wave method[46,47]. A plane-wave energy cut-off of 520 eV and a $20 \times 20 \times 1$ $k$-mesh were employed. The generalized gradient approximation (GGA) with the Perdew-Burke-Ernzerhof (PBE) functional[48] was used for structure optimization of $HfTe_2$. Freestanding films were modeled with a 16-Å vacuum gap between adjacent layers in the supercell. Experimental in-plane lattice constants were used in calculations ($a = 3.967$ Å for 1 TL and 3.95 Å for multilayer films)[22,31]. For the 2 TL and 3 TL, the van der Waals dispersion energy correction term of DFT-D3 type[49] was used to optimize structures. Spin-orbit couplings are included in the calculations. Phonon calculations were carried out using the supercell method as implemented in the Phonopy package[50]. Phonon frequencies are well converged at the $\bar{\Gamma}$ and $\bar{M}$ points for a 4 x 4 supercell. We also checked the convergence against planewave energy cut up to 620 eV and the k-point sampling up to a 48 x 48 k-grid and find no sign of imaginary phonons. Topological invariants are computed by tracking the evolution of hybrid Wannier functions implemented in the Z2pack[51,52].

**Film structure characterization**

Extended Data Fig. 1a shows angle-integrated core level scans of the characteristic peaks of Hf and Te for the 1T phase $HfTe_2$. Sharp RHEED patterns (Extended Data Fig. 1c) reveal a high-crystalline-quality and well-ordered single-layer $HfTe_2$ grown by MBE on HOPG. Multilayer films were grown layer-by-layer as RHEED intensity of the diffraction spots oscillates during the growth (Extended Data Fig. 1b), which is used to control and determine the layer thickness of the films. Formation of the multilayer of $HfTe_2$ is also evidenced by evolution of the band structure measured by ARPES (Fig. 1). Extended Data Fig. 1d shows the evolution of RHEED patterns during a 3 TL film growth. The similar patterns for $N$-layer films ($N$ = 1 - 3) demonstrates the in-plane lattice constants show little change with thickness. The in-plane lattice constants are determined as a = 3.96 ± 0.04 Å, consistent with the previous reports[22,31].

**Temperature dependence of the band structure and Fermi surface maps**

Extended Data Figs. 2a and 2b show systematic scans of the bands on 1 TL $HfTe_2$/BLG/SiC ($n$ - $p$ = -5.4 x $10^{12}$/$cm^2$) taken with 45 and 60 eV photons along $\overline{\Gamma M}$ as a function of temperature, revealing details of the development of the flat band feature, opening of the gap, and the intensity variation of the folded bands in connection with the metal-insulator transition. The flat valence band top becomes more prominent in spectra taken with 60 eV, which reveals that the intensity of bands is affected by the cross section of the photoelectron. We note the conduction band edge emerges at the low temperatures and the full width at half maximum (FWHM) of the conduction band peak is 33 meV at 10 K, only a third of the value of the valence band peak (108 meV). This result suggests that only part of the conduction band crosses over the Fermi level and the band gap determined in the main text is underestimated.

**Calculated band structures and topological properties of $N$-layer HfTe$_2$ ($N$ = 1 - 3)**

The computed band structures of ($1 \times 1$) lattice for $N$-layer HfTe$_2$ in the normal phase are shown in Extended Data Fig. 6. Of note is that DFT results predict a larger overlap between the valence band top and the conduction band bottom than the experiment in the normal phase due to the inherent limitations of the DFT method. For 2 and 3 TL HfTe$_2$, we note that there is a band inversion between Hf $5d$ and Te $5p$ states at the $\bar{\Gamma}$ point around 0.6 eV above the Fermi level in the normal phase, similar to the previous report and the bulk case with MBJ calculations, which suggests a topological nontrivial properties in low dimensional HfTe$_2$[21-24,53]. Furthermore, we calculate the hybrid Wannier charge center (HWCC) evolution along the $\overline{\Gamma M}$ direction, as presented in Extended Data Fig. 7. The topological nontrivial character is demonstrated by a crossing between the middle of the largest gap and the HWCC near the $\bar{\Gamma}$ point. This result makes HfTe$_2$ an exciting platform to study the interplay between excitonic and topological orders[16].

**Data availability**

The data shown in the figures and other findings of this study are available from the corresponding authors upon reasonable request.

**Methods references**

**Acknowledgments** We thank Prof. Tai C. Chiang, Prof. Wei Ku and Prof. Xiaoyan Xu for helpful discussions. The work at Shanghai Jiao Tong University is supported by the Ministry of Science and Technology of China under Grant Nos. 2021YFE0194100, 2022YFA1402400, 2021YFA1401400, and 2021YFA1400100, the Science and Technology Commission of Shanghai Municipality under Grant No. 21JC1403000, National Natural Science Foundation of China (Grant Nos. 12174250, 12141404), Shanghai Jiao Tong University 2030 Initiative. Y. H. C. acknowledges support by the Ministry of Science and Technology, National Center for Theoretical Sciences (Grant No. 110-2124-M-002-012) and National Center for High-performance Computing in Taiwan. Part of this research is supported by ME2 project under Contract No. 11227902 from National Natural Science Foundation of China.

**Author contributions** P.C. conceived the project. Q.G., P.C. with the aid of H.Y.C., Y.C.Y, Z.T.L., and D.W.S. performed MBE growth, ARPES measurements, and data analysis. Y.H.C. and K.T. performed calculations. P.F.J and S.W.J. performed Raman measurements. P.C., Q. G., H.Y.C.,S.S.Y., and X.L.L. carried out x-ray diffraction measurements. P.C., Y.H.C., and Q.G. interpreted the results. P.C. wrote the paper with input from other coauthors.

**Competing interests** Authors declare no competing interests.

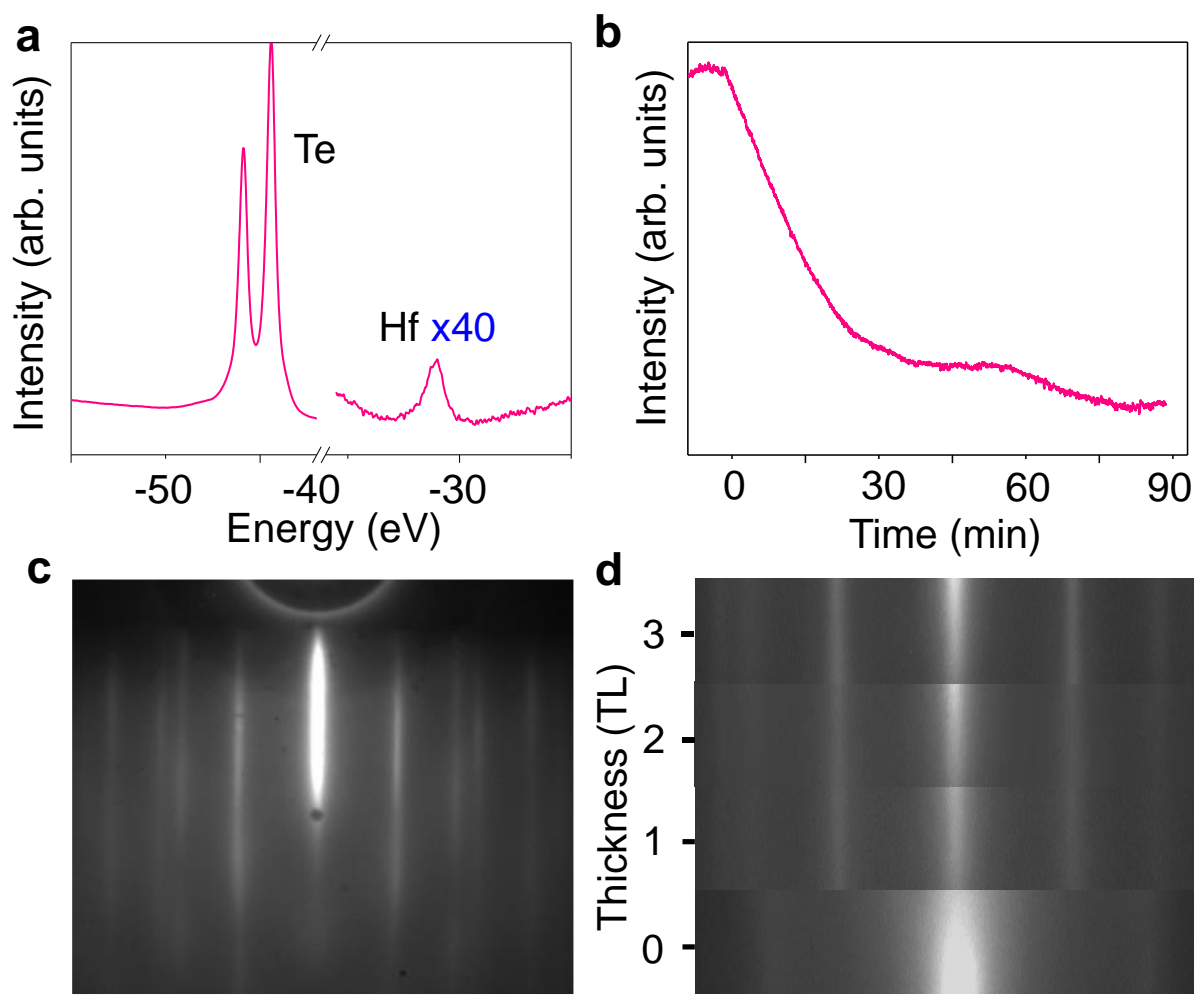

**Extended Data Fig. 1 | Core-level scans and RHEED patterns on HfTe₂ Films.** (**a**) Core-level photoemission spectra taken with 90 eV photons for a 1 TL HfTe$_2$/BLG/SiC film. (**b**) An example of RHEED intensity oscillations of the central diffraction spot as a function of time. (**c**) A RHEED pattern of single-layer HfTe$_2$ film taken at room temperature. (**d**) RHEED patterns as a function of the HfTe$_2$ thickness, indicating similar in-plane lattice constants in single-layer and multilayer films.

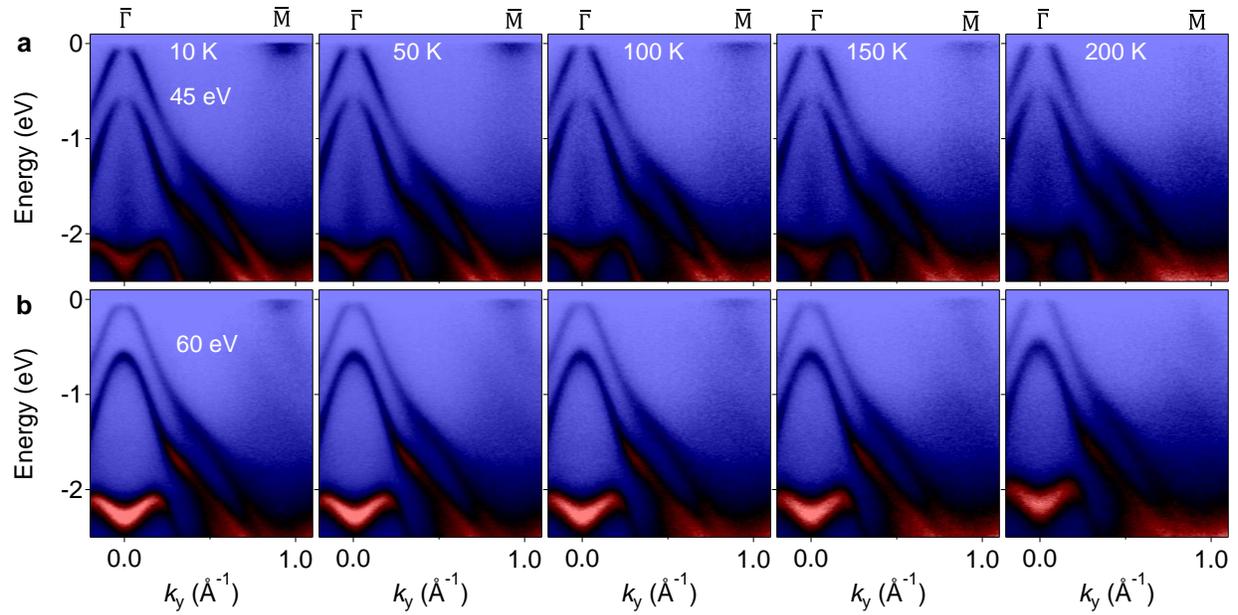

**Extended Data Fig. 2 | Temperature dependence of the band structure of single-layer HfTe₂.**
ARPES maps on 1 TL HfTe$_2$/BLG/SiC ($n$ - $p$ = -5.4 x 10$^{12}$/cm$^2$) taken with (**a**) 45 and (**b**) 60 eV
photons along the $\overline{\Gamma M}$ direction at selected temperatures.

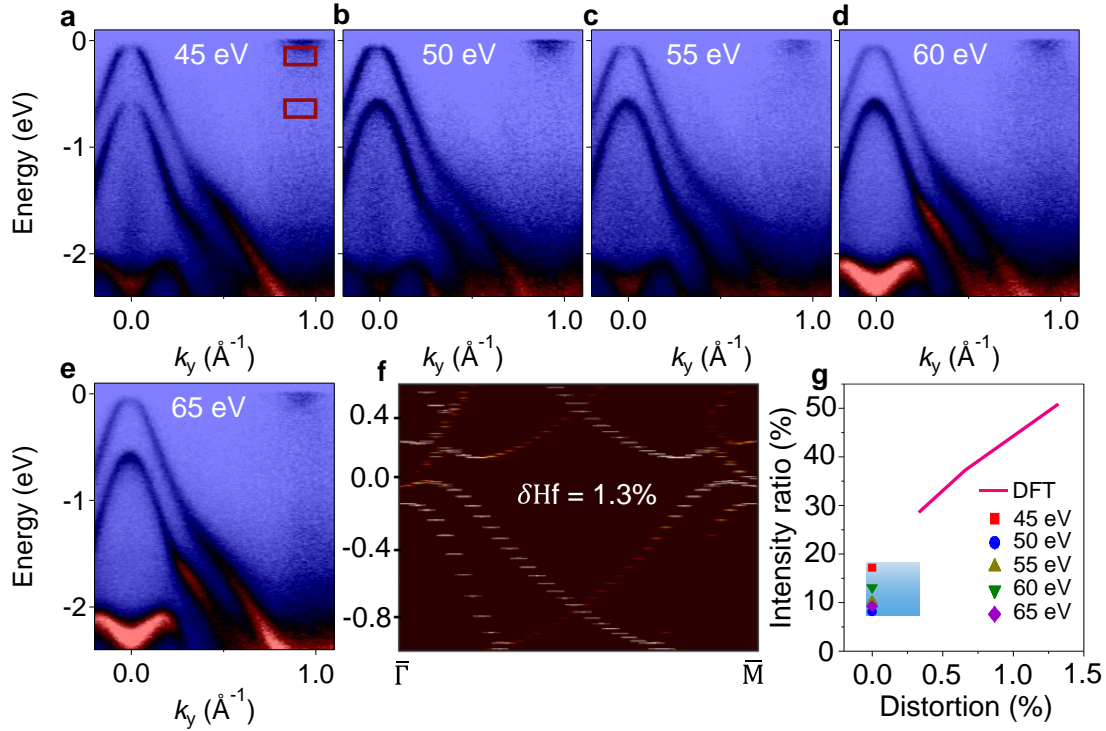

**Extended Data Fig. 3 | Photon energy dependence of the ARPES spectra for 1 TL HfTe₂.** (**a**)-(**e**) ARPES maps taken with various photon energies along $\overline{\Gamma M}$ for 1 TL HfTe₂/BLG/SiC ($n$ - $p$ = -5.4 x $10^{12}$/cm²) at 10 K. (**f**) Calculated unfolded band structure for a (2 × 2) superstructure using the PBE method with a imposed Hf displacement of 1.3% of the lattice constant. (**g**) Relative intensity ratio of the lower folded valence band (at ~-0.5 eV) to the upper folded band with background subtracted. Regions of interest used for integrating the folded band intensity were indicated as red boxes shown in panel (**a**). To reduce the matrix element effect, we extract the relative folded band intensities ($I_f$) with respect to the main band intensities ($I_m$). The DFT calculated ratio results with a series of imposed lattice distortions are shown as a solid line. The smallest imposed lattice distortion is 0.33% of the lattice constant since the calculated two folded bands are indistinguishable in energy below this value. It is clear that the experimental results with

different photon energies are well below the smallest predicted value, resulting in an upper bound for the magnitude of the lattice distortion (~0.013 Å) in 1 TL $HfTe_2$.

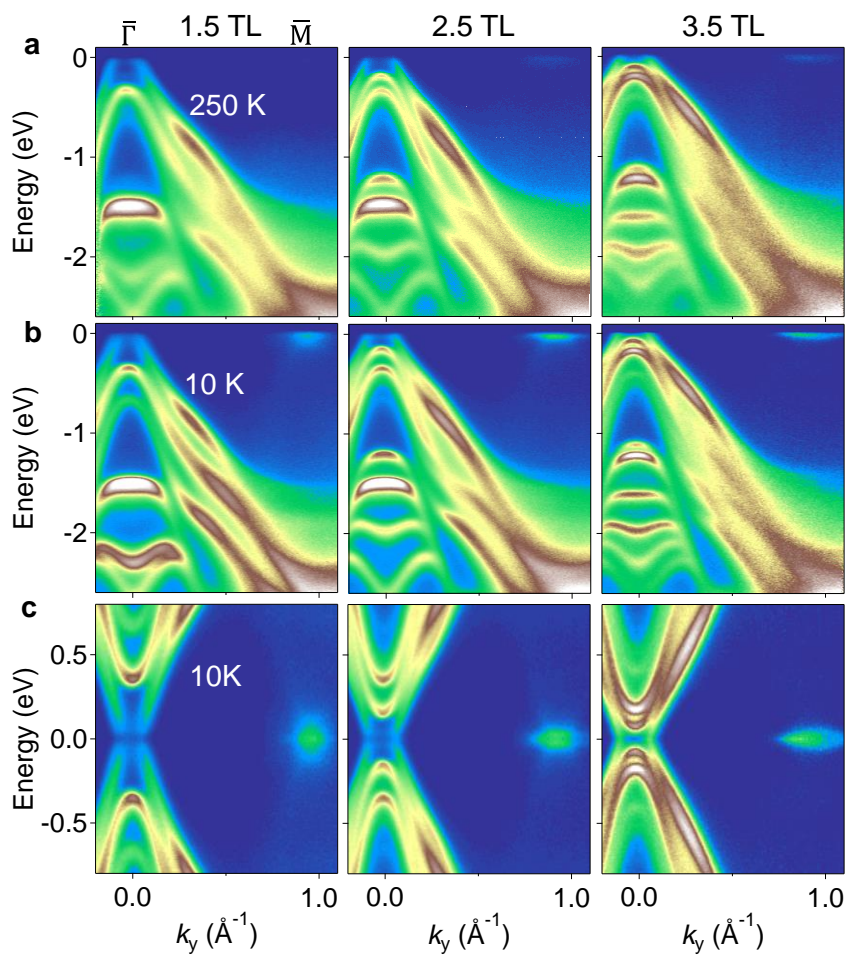

**Extended Data Fig. 4 | More thickness dependent band structures of low dimensional HfTe₂.** ARPES maps along $\overline{\Gamma M}$ for low dimensional HfTe₂ at (**a**) 250 K and (**b**) 10 K. (**c**) Corresponding symmetrized ARPES maps in energy about the Fermi level at 10 K. The decimals in the layer thickness are nominal and determined from the relative intensity of the subbands in the adjacent layers.

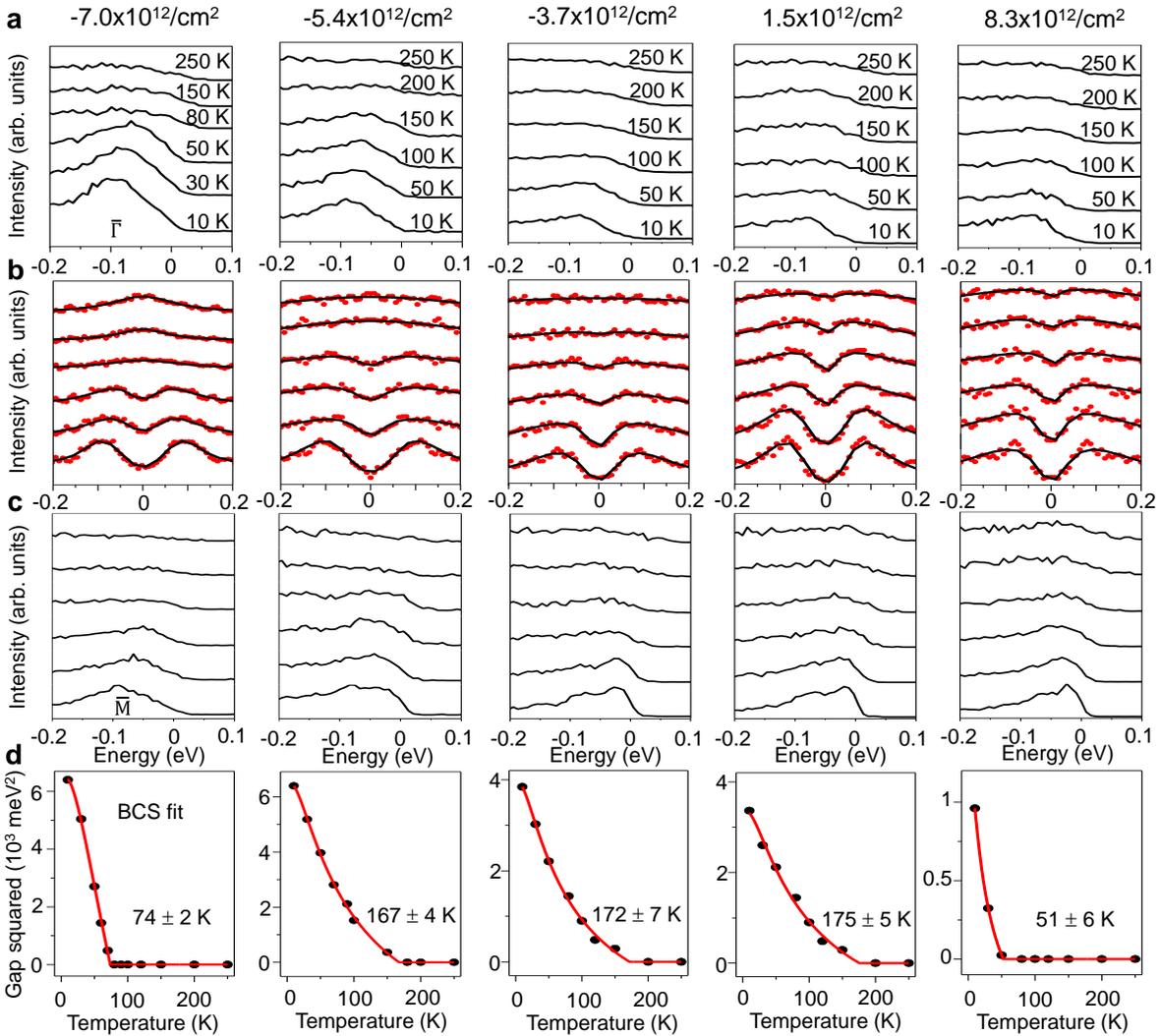

**Extended Data Fig. 5 | Temperature dependence of the band structure and the energy gaps for 1 TL HfTe₂ with different doping levels**. (**a**) EDCs at the zone center at selected temperatures between 10 and 250 K. (**b**) Corresponding symmetrized EDCs. The temperature dependence of the Fermi-Dirac distribution can be canceled out by symmetrization. The fits to a phenomenological BCS-type function are indicated as solid curves. (**c**) EDCs at the $\bar{\text{M}}$ point at selected temperatures between 10 and 250 K. (**d**) The extracted temperature dependence of the square of the energy gap. The red curves are fitting results using the BCS-type mean-field equation.

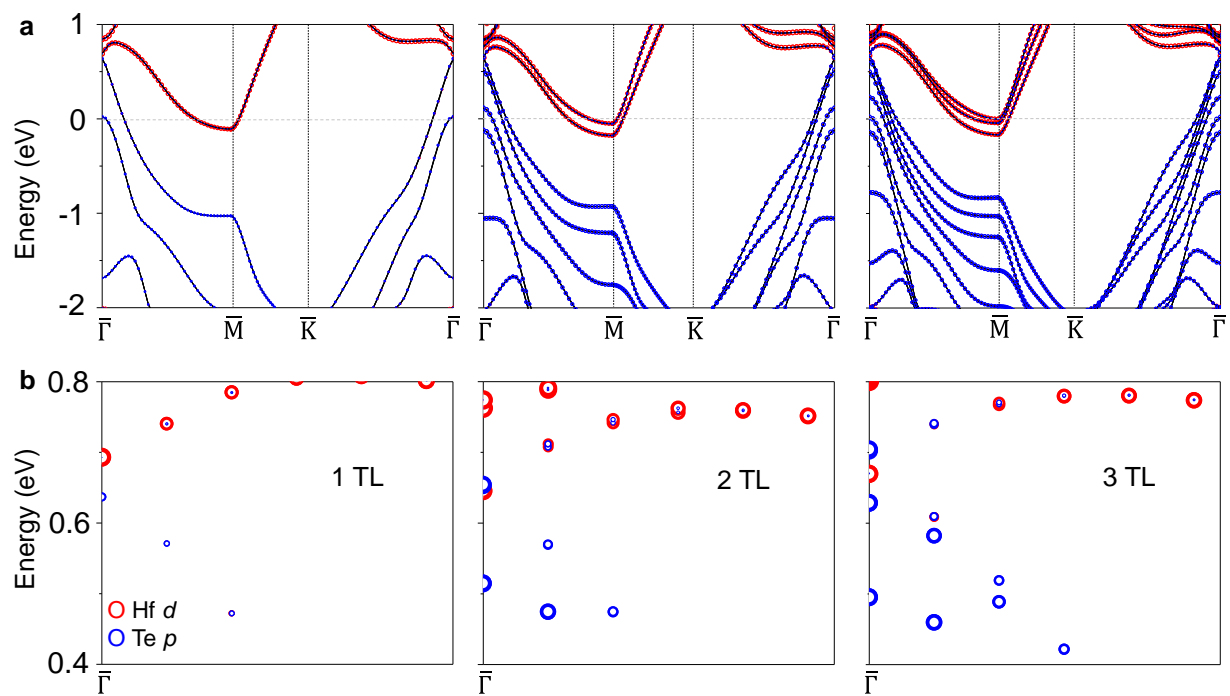

**Extended Data Fig. 6 | DFT band dispersions of *N*-layer HfTe₂ (*N* = 1-3).** (**a**) Calculated band structures in the normal phase for *N*-layer HfTe₂. (**b**) Enlarged plots to show details around the zone center. The band characters, Hf 5*d* or Te 5*p*, are color coded.

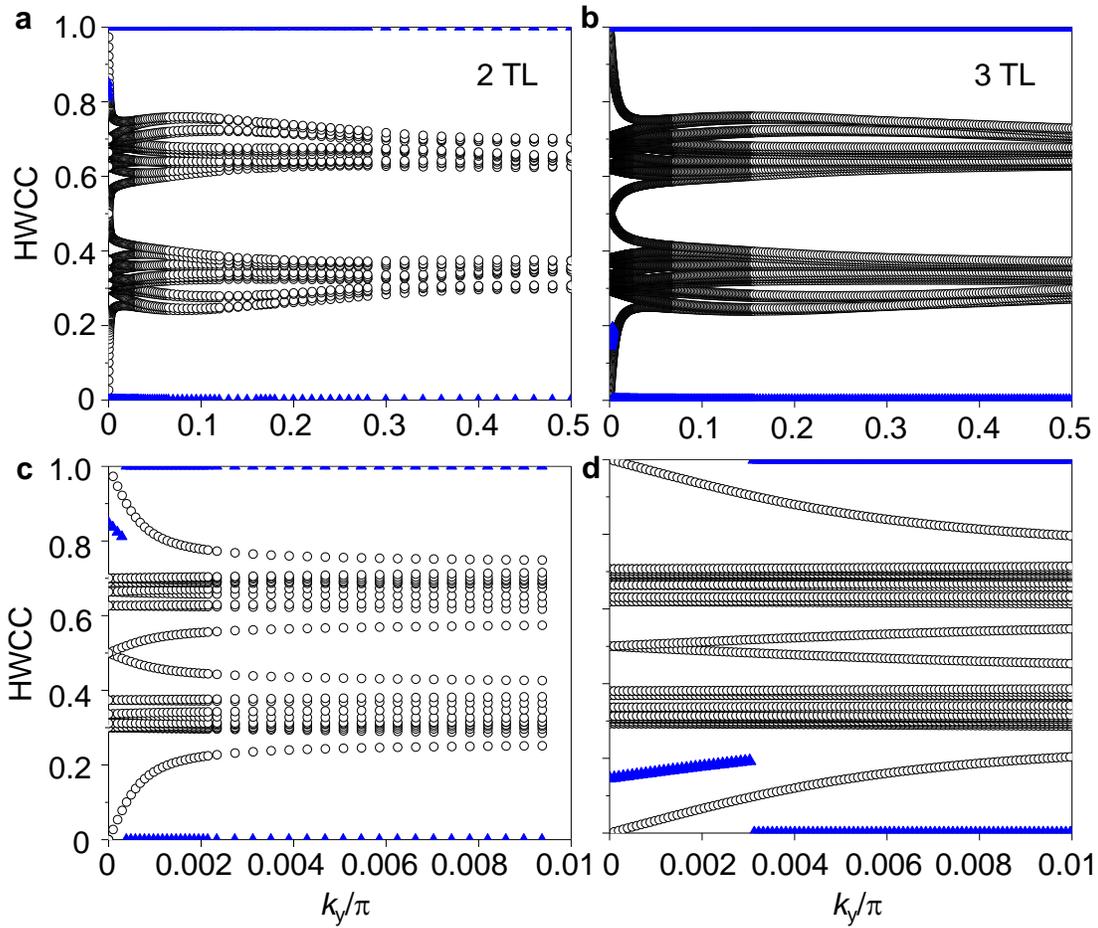

**Extended Data Fig. 7 | Topological Z2 invariant calculations**. Evolution of HWCC (black empty dots) and the middle of the larges gap (blue triangles) along the $\overline{\Gamma M}$ direction for (**a**) 2 TL and (**b**) 3 TL HfTe$_2$. (**c**) and (**d**) are the corresponding zoom-in details near the $\overline{\Gamma}$ point.

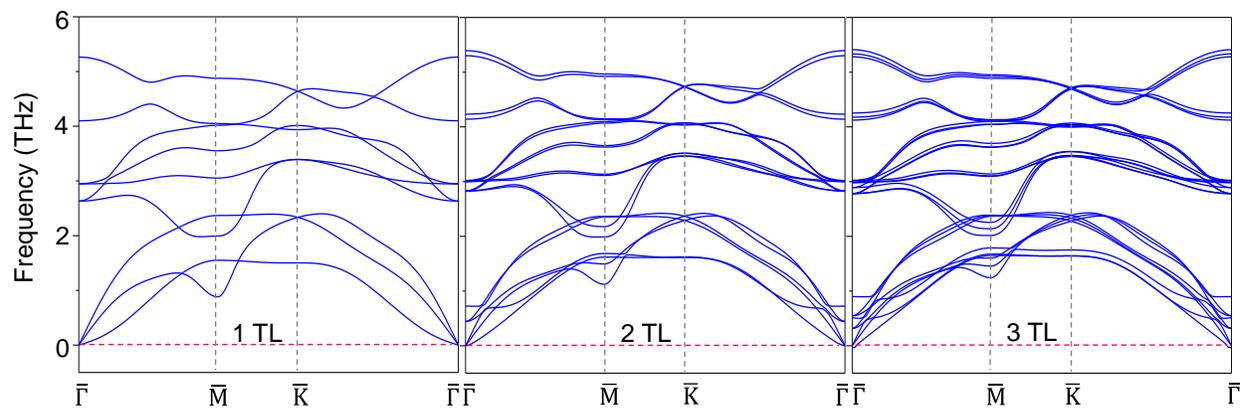

**Extended Data Fig. 8 | Calculated phonon dispersions**. Phonon dispersions for $N$-layer HfTe$_2$ ($N$ = 1, 2, 3) in the normal phase computed from first principles.

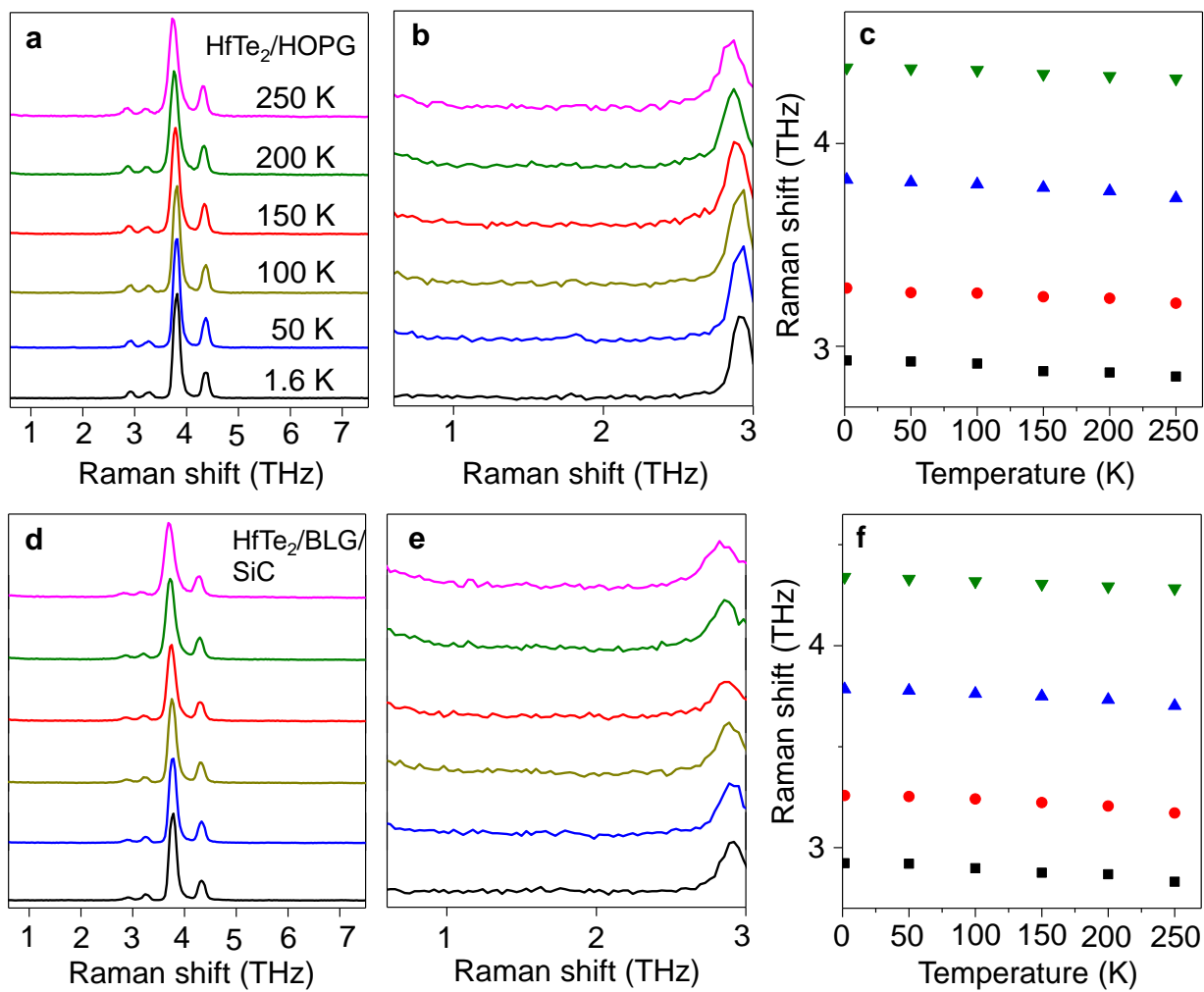

**Extended Data Fig. 9 | Raman scattering spectra of single layer HfTe₂ films**. Temperature-dependent Raman scattering spectra for 1 TL HfTe₂ on (**a**) HOPG and (**d**) bilayer-graphene-terminated SiC. (**b**), (**e**) zoom-in of the Raman spectra in a region below 3 THz. No new peaks were observed across the measured temperature range that could indicate any structural distortion. (**c**), (**f**) Corresponding extracted Raman frequencies of the E and A1 modes between 2.8 and 4.3 THz as a function of temperature.

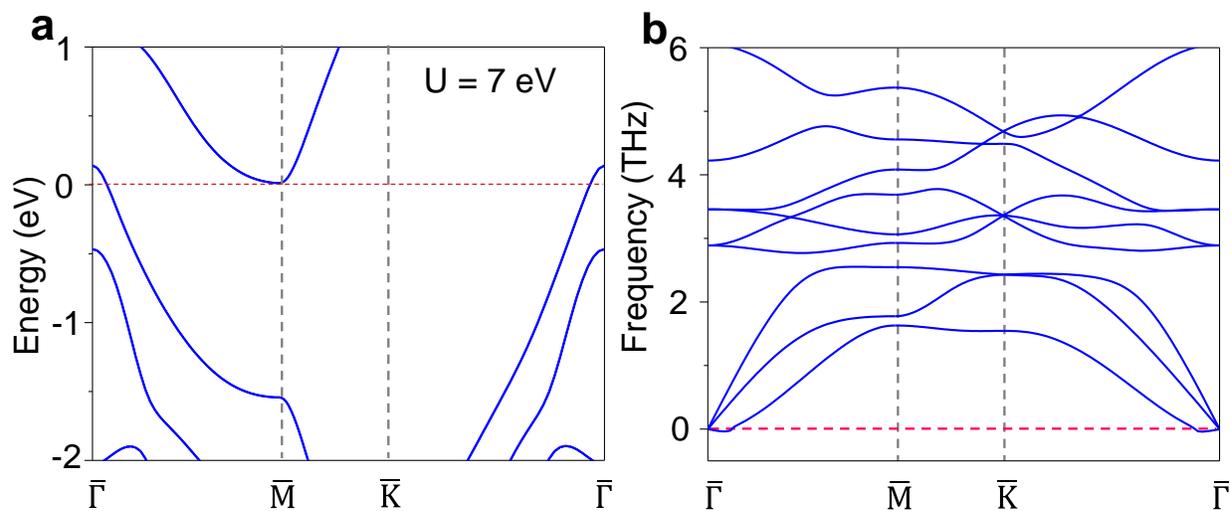

**Extended Data Fig. 10 | Calculated band structure and phonon dispersions with PBE + U = 7 eV**. Hubbard U is included to account for the electron localization and (**a**) the band structure agrees well with the experiments. (**b**) The calculated phonon dispersions with U also shows no imaginary frequencies, predicting no structural instability.

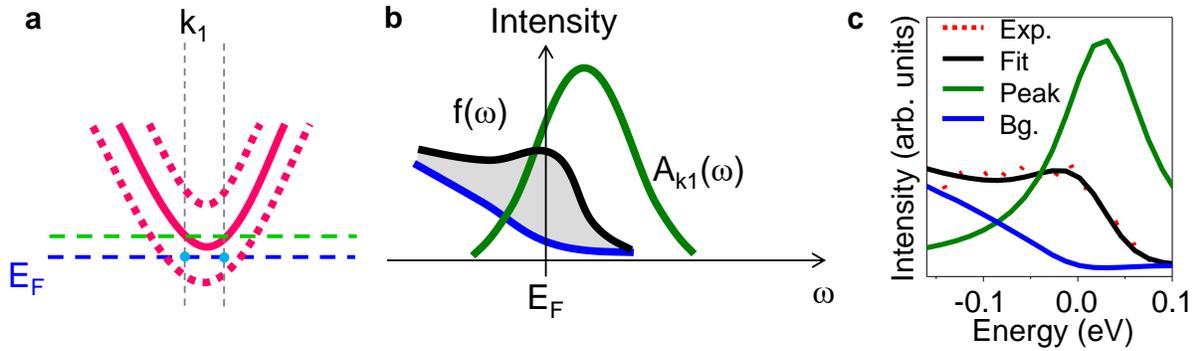

**Extended Data Fig. 11 | Carrier density estimation for the conduction band tail**. (**a**) A schematic diagram of the conduction band spectral function. The magenta solid line indicates the center of mass of the conduction band and the magenta dashed lines demonstrate the broadened spectral function. The cyan dots represent the peak position of the MDC fitting results used to calculate the carrier density, which actually means the carrier density at the position as if the Fermi level is at the green dashed line. (**b**) Schematic of the EDC along the cut through one of the cyan dots. The green line is the conduction band peak at this position. As a simple approximation, the ratio of the shadow area to that of the green line represents the ratio of the carrier density at Fermi level ($E_F$) to that at the position marked by the green line. Since the carrier density at the green line position was obtained by formula $N = g_v \frac{k_F^2}{2\pi}$, we can thus estimate the carrier density at $E_F$. (**c**) An example fit for the highest electron doping case. The black curve is the fitting result which is a combination of the Shirley background, the Lorentzian peak multiplied by the Fermi-Dirac function and convoluted with a Gaussian corresponding to the experimental energy resolution.